\documentclass[aps,prl,twocolumn,groupedaddress,nofootinbib]{revtex4-1}

\usepackage{amsmath,amssymb,MnSymbol}
\usepackage{graphicx}

\begin{document}
\def\query#1{\marginpar{\begin{flushleft}\footnotesize#1\end{flushleft}}}
\newcommand{\br}{\bra{p',s'}}
\newcommand{\kt}{\ket{p,s}}
\newcommand{\p}{\partial_}
\newcommand{\m}{\mathbf}{}
\newcommand{\bs}{\boldsymbol}{}
\newcommand{\tf}{\tau_1}
\newcommand{\ts}{\tau_2}
\newcommand{\eq}{\begin{eqnarray}}
\newcommand{\en}{\end{eqnarray}}

\title{Nucleon in a periodic magnetic field}


\author{Andria Agadjanov$^{a}$, Ulf-G. Mei{\ss}ner$^{a,b}$ and Akaki Rusetsky$^a$}
\affiliation{$^a$Helmholtz-Institut f\"ur Strahlen- und Kernphysik (Theorie) and
Bethe Center for Theoretical Physics, Universit\"at Bonn, D-53115 Bonn, Germany}
\affiliation{$^b$Institute for Advanced Simulation (IAS-4), Institut f\"ur Kernphysik 
	(IKP-3) and  J\"ulich Center for Hadron Physics,
	Forschungszentrum J\"ulich, D-52425 J\"ulich, Germany}


\date{\today}

\begin{abstract}
The energy shift of a nucleon in a static periodic magnetic field is
 evaluated at second
order in the external field strength in perturbation theory. It is shown that
the measurement of this energy shift on the lattice allows one to determine the 
unknown subtraction function in the forward doubly virtual Compton scattering amplitude. 
The limits of applicability of the obtained formula for the energy shift are discussed.
\end{abstract}


\maketitle

\section{Introduction}

The doubly virtual Compton scattering process has several important 
phenomenological implications at low energies. The analysis of the 
proton-neutron mass difference\footnote{Note that, recently, a substantial progress 
has been achieved in the direct evaluation of this difference on the 
lattice~\cite{Borsanyi:2014jba,Horsley:2015eaa}. 
A comparison of different approaches provides constaints 
on the behavior of the Compton amplitudes.} 
\cite{Gasser:1974wd,Gasser:2015dwa,WalkerLoud:2012bg,Erben:2014hza} 
relies on the knowledge of the relevant spin-independent invariant amplitudes $T_1$ and $T_2$. 
The same amplitudes appear in the study of the Lamb shift in the muonic hydrogen (see, e.g., 
Refs.~\cite{Birse:2012eb,Pachucki:1999zza,Peset:2014jxa}). 

The experimental data on the structure functions completely determine the amplitude 
$T_2$. However, they do not fix the subtraction function $S_1$ in the dispersion relation 
for the amplitude $T_1$. This function depends on the photon virtuality $q^2$. The 
low-energy theorem establishes a relation between the value of the $S_1(q^2)$ at 
$q^2=0$ and the magnetic polarizability of the nucleon.  Further,
the asymptotic behavior of this function for large spacelike $q^2$ is fixed by the operator
product expansion in QCD~\cite{Collins:1978hi,Erben:2014hza,Hill:2016bjv}. The behavior of the 
function at intermediate values of $q^2$ is, however, completely arbitrary.
Earlier calculations of the forward Compton scattering
amplitude within chiral effective field theories at low photon virtualities
have been carried out in Refs.~ \cite{Bernard:1991ru,Bernard:1992qa,Ji:1999sv,Bernard:2002pw}. 
The latest calculations of the quantity $S_1$ are contained in 
Refs.~\cite{Birse:2012eb,Hill:2011wy,Alarcon:2013cba,Lensky:2014dda,Nevado:2007dd,Peset:2014jxa}.
The convergence of the chiral expansion is, however, questionable even at very 
low $q^2$. Further, several authors have used phenomenological parametrizations of the function 
$S_1$ in their calculations~\cite{Pachucki:1999zza,WalkerLoud:2012bg,Erben:2014hza}.
Unfortunately, this introduces a systematic uncertainty in the calculated observables that 
is very hard to control.

An interesting possibility to determine the subtraction function $S_1$ was discussed in
Refs.~\cite{Gasser:1974wd,Gasser:2015dwa}. If the forward scattering amplitude does
not contain any so-called fixed pole (this issue is related to the so-called
Reggeon dominance hypothesis), then $S_1(q^2)$ for all $q^2<0$ is uniquely 
determined by the dispersion integral over the
electroproduction cross sections in the physical region. Therefore,
if one could calculate the function $S_1(q^2)$ directly and compare with the result
obtained by using the Reggeon dominance hypothesis, in principle 
one would be able to answer a question, whether  a fixed pole is present 
in the forward Compton 
amplitude or not (for the recent phenomenological evaluation of $S_1$, see
Refs.~\cite{Gasser:2015dwa,Gorchtein:2013yga,Tomalak:2015hva}). For instance, note that 
the universality hypothesis,
stated in Ref.~\cite{Brodsky:2008qu}, does not exclude the presence of a 
fixed pole in $T_1$.
Further, a calculation of $S_1(q^2)$
 would allow one to evaluate the proton-neutron electromagnetic mass shift
and the two-photon exchange contribution to the muonic hydrogen Lamb shift in a manner
devoid of any model dependence.
Hence, there is strong  interest  in a  direct calculation  of the function $S_1(q^2)$.

At present, lattice QCD is the only first-principle approach capable of handling
the above problem. There are two ways to determine $S_1$. In the first method, one 
directly calculates the four-point function that describes  Compton scattering. 
This  is a straightforward but computationally very demanding task. Until now, this 
approach has been used in the computation of light-by-light scattering~\cite{Green:2015sra}, and 
also in the study of the long-distance effects in rare kaon 
decays~\cite{Christ:2016eae,Christ:2016mmq}. 
An alternative method is based on the observation that the Compton scattering 
amplitude can be inferred from the behavior of the nucleon  two-point function in the 
presence of a weak external electromagnetic field.
In recent years, the external field method has become a powerful tool to study the 
electromagnetic properties of the nucleon and light nuclei. In particular, it has been 
used for the case of a constant and uniform magnetic field \cite{Tiburzi:2012ks,Chang:2015qxa,Parreno:2016fwu}. 
Such a field configuration allows one to determine the magnetic moments and magnetic 
polarizabilities through the extracted energy shift induced by the magnetic field. By 
applying nonuniform and time-dependent fields, one can determine spin polarizabilities 
as well (see, e.g., Refs. \cite{Detmold:2006vu,Davoudi:2015zda}). 
Hence, the external field approach for the calculation of 
$S_1$ should be feasible in practice.

In this paper, we demonstrate that measuring the energy shift of a single nucleon
state in a  {\it static periodic} magnetic field on the lattice enables one to determine the function
$S_1(q^2)$ at nonzero values of $q^2$. Unlike a  constant magnetic field, which 
generates the harmonic oscillator potential and Landau levels, in the present case the 
spectrum and the eigenstates of the Hamiltonian  can not be obtained analytically. 
Nevertheless, the energy is still conserved in the static magnetic field and, as long as the 
potential remains ``small,''  perturbation theory can be applied to the free energy 
spectrum (there is a similar approach in solid-state physics, which is called the nearly 
free electron model, and more generally, the empty lattice approximation~\cite{Kittel}). 
We show that, at second order in the magnetic field strength, the energy shift of the
one-nucleon ground state is proportional to the quantity $S_1(q^2)$. The limits of
applicability of this perturbative expression are also discussed.

\section{Compton scattering}
Let us start with the basic definitions. The Compton scattering amplitude is given by
\begin{equation}\label{Compton tensor}
T^{\mu\nu}(p',s';p,s;q)=\frac{i}{2}\int d^4x e^{i q\cdot x}\langle p',s'|Tj^\mu(x)j^\nu(0)|p,s\rangle\, .
\end{equation}
Here, $j^\mu(x)$ denotes the electromagnetic current, $q$ is the four-momentum of the final photon, 
$p\,(p')$ and $s\,(s')$ are the four-momenta and spins
 of the initial (final) nucleon, respectively. 
Considering  forward scattering $p'=p$ and performing  the spin-averaging in 
Eq. (\ref{Compton tensor}), one arrives at
\eq\label{Tensor_averaged}
T^{\mu\nu}(p,q)=\frac{1}{2}\sum_s T^{\mu\nu}(p,s;p,s;q).
\en
The tensor $T^{\mu\nu}(p,q)$  is related to the invariant amplitudes $T_1,\,T_2$ through the expression 
(see, e.g., Refs. \cite{Gasser:2015dwa,Tarrach,Bernabeu:1976jq}):
\eq\label{Taveraged}
T^{\mu\nu}(p,q)= T_1(\nu,q^2)K_1^{\mu\nu} +T_2(\nu,q^2)K_2^{\mu\nu},
\en
where the kinematic structures $K_1^{\mu\nu},\, K_2^{\mu\nu}$ read
\eq
K_1^{\mu\nu}&=& q^\mu q^\nu-g^{\mu\nu}q^2,\nonumber\\[2mm]
K_2^{\mu\nu}&=& \frac{1}{m^2}\Big\{(p^\mu q^\nu+p^\nu q^\mu)p\cdot q
-g^{\mu\nu}(p\cdot q)^2-p^\mu p^\nu q^2\Big\}\, .\nonumber\\
\en
Here, $m$ is the nucleon mass and $\nu\equiv p\cdot q/m$. The subtraction function $S_1$ is defined as
\eq
S_1(q^2)=T_1(0,q^2)\, .
\en
The quantity $S_1(q ^2)$ can be split into the elastic and inelastic parts
(see, e.g., Ref.~\cite{Gasser:2015dwa}). The elastic part is singular
at $q ^2\to 0$, whereas the inelastic part
is regular and is related to the nucleon magnetic polarizability at $q ^2=0$.
Note that the quantity $S_1(q^2)$ is real, because in this kinematical region there are no
multi-particle singularities. Indeed,  introducing the Mandelstam variable 
$s=(p+q)^2=m^2+2m\nu+q^2$, it is immediately seen that, for $\nu=0$ and 
$q^2<0$, we have $s<m^2$, meaning that one is below the inelastic threshold.

It is well known that the Compton tensor given in Eq.~(\ref{Compton tensor}) can be
obtained by  expanding the two-point function of a nucleon in an external electromagnetic
field to  second order. Introducing the notation  $A_\mu(x)$ for the external 
potential, it is actually seen that the nucleon propagator can be
expanded as follows:
\begin{widetext}
\eq\label{eq:PsiPsi}
\langle 0|T\Psi(x)\bar\Psi(y)|0\rangle_A
&=&\langle 0|T\Psi(x)\bar\Psi(y)|0\rangle_0
+\frac{i}{1!}\int d^4zA_\mu(z)\langle 0|T\Psi(x)\bar\Psi(y)j^\mu(z)|0\rangle_0
\nonumber\\[2mm]
&+&\frac{i^2}{2!}\int d^4zd^4vA_\mu(z)A_\nu(v)\langle 0|T\Psi(x)\bar\Psi(y)j^\mu(z)j^\nu(v)|0\rangle_0+\cdots\, .
\en
\end{widetext}
Here, $\Psi(x)$ denotes the (composite) nucleon field operator in QCD and the subscript
``0'' refers to the quantities evaluated in QCD without any external field.
Note that Eq.~(\ref{eq:PsiPsi}) 
is written down for  connected matrix elements 
(the subscript ``conn'' is omitted everywhere for brevity).
Further, performing the Fourier transform in 
Eq.~(\ref{eq:PsiPsi}), amputating the external nucleon legs, and putting
external nucleons on the mass shell,
we see that the nucleon electromagnetic vertex $\langle p',s'|j^\mu(0)|p,s\rangle$
emerges at order $A$. At order $A^2$, as already
mentioned, the scattering amplitude given in Eq.~(\ref{Compton tensor}) is obtained
from the matrix element 
$\langle 0|T\Psi(x)\bar\Psi(y)j^\mu(z)j^\nu(v)|0\rangle_0 $, and
so on.

On the other hand, since one is below the inelastic
threshold, one may describe the nucleon two-point function within the nonrelativistic
effective field theory as well, matching the couplings of the effective Lagrangian to
the pertinent expressions in QCD. The advantage of this approach will become apparent,
when the energy spectrum of a system in a finite box will be considered, as in the 
nonrelativistic effective theory, the energy levels are obtained by merely solving the 
Schr\"odinger equation. 

In the present work we consider the matching and the subsequent calculation of the energy
shift in a very condensed manner. A detailed treatment of these issues, as well
as a thorough study of the finite-volume spectrum of the Hamiltonian with periodic
potentials wil be the subject of a separate publication~\cite{future}. In brief, the 
procedure looks as follows. At the first stage, the matching of the relativistic and 
nonrelativistic theories is carried out in the infinite volume. The matching ensures
that the on-shell coefficients in the expansion of the two-point function 
up to and including $O(A^2)$ are reproduced in the nonrelativistic theory.
At the next step, one uses the nonrelativistic Hamiltonian, whose couplings are
fixed through the matching, to calculate the energy levels in a  finite volume. Note that
this procedure is self-consistent, since the couplings of the 
effective Hamiltonian encode solely the short-range physics that does not get
altered by placing the system in a large box.

Above, we have already written down the expansion of the nucleon two-point 
function in the external field in QCD. Now, we want to do the same 
in the effective theory. The corresponding Lagrangian, which describes 
the gauge-invariant interaction of the nucleon with an external electromagnetic field, 
has the following general form:
\eq\label{L_eff_A}
{\cal L}_{\rm eff}={\cal L}_0+{\cal L}_1+{\cal L}_2+\cdots.
\en
Here, ${\cal L}_0$ is the free nucleon Lagrangian (we remind the reader that we are below the inelastic cuts)
\eq
{\cal L}_0=\psi^\dagger2{W}(i\p t-{W})\psi,\quad {W}=\sqrt{m^2-\bs\nabla^2}\, .
\en
In the above expression, $\psi(x)$ is a two-component field that describes the nonrelativistic nucleon. 
It is seen that the relativistic dispersion relation for the energy of the free nucleon with the
three-momentum $\m p$, $w(\m p)=\sqrt{m^2+\m p^2}$, is satisfied. Also, the factor $2W$ ensures that 
the one-particle states have the relativistic normalization 
(see, e.g., Refs.~\cite{Colangelo:2006va,Gasser:2011ju}).

Further,  ${\cal L}_1$ is linear in the external
field $A_\mu$,  ${\cal L}_2$ is quadratic in $A_\mu$, and so on. 
Note that ${\cal L}_1$ and ${\cal L}_2$ should
contain an infinite set of operators with an arbitrary number of  space derivatives, 
which act on the fermion and external fields. This differs from the situation for vanishing
$q^2$ (the pertinent Lagrangian is given, e.g., in Ref.~\cite{Detmold:2006vu}).
However, as we shall see, the infinite number of terms will effectively sum up in a single
function. 

Applying the equation of motion to eliminate the time derivatives, and performing partial 
integration, ${\cal L}_1$ and ${\cal L}_2$ can be brought into the form
\begin{widetext}
\eq \label{L1L2}\nonumber
{\cal L}_1&=&\sum_{m,n=0}\,A_\mu(x)[\p{i_1}\dots \p{i_n}\psi_{s'}^\dagger(x)]\Gamma_{s's}^{i_1\dots i_n,\,j_1\dots j_m,\,\mu}[\p{j_1}\dots \p{j_m}\psi_{s}(x)]\, ,\\
\vspace{0.2cm}
{\cal L}_2&=&\sum_{l,m,n=0}\,A_\nu(x)[\p{\mu_1}\dots \p{\mu_l}A_\mu(x)]\,[\p{i_1}\dots \p{i_n}\psi_{s'}^\dagger(x)]\Pi_{s's}^{i_1\dots i_n,\,j_1\dots j_m,\,{\mu_1}\dots{\mu_l},\,\mu\nu}
[\p{j_1}\dots \p{j_m}\psi_{s}(x)],
\en
\end{widetext}
where $\Gamma_{s's}^{i_1\dots i_n,\,j_1\dots j_m,\,\mu}$ and $\Pi_{s's}^{i_1\dots i_n,\,j_1\dots j_m,\,{\mu_1}\dots{\mu_l},\,\mu\nu}$ denote  low-energy constants. 
The Latin indices run from 1 to 3 (only space derivatives), whereas the Greek
indices run from 0 to 3. 
The derivatives in the square brackets act only on the function within the brackets. Also, as a convention, 
the values $m,n=0$ correspond to no derivatives in Eq.~(\ref{L1L2}). 

The comparison of the two-point functions  at $O(A)$,
 calculated in the different theories, leads to the  matching condition
\begin{widetext}
\eq\label{mA}
 \sum_{m,n=0}\,(-ip')_{i_1}\dots (-ip')_{i_n}(ip)_{j_1}\dots (ip)_{j_m}\Gamma_{s's}^{i_1\dots i_n,\,j_1\dots j_m,\,\mu} = \left\langle p',s'| j^\mu(0)|p,s\right\rangle\, .
\en
\end{widetext}
The expression of the second derivative of the two-point function with respect to the
external field in the effective theory consists of two parts: the contribution from 
${\cal L}_2$ and the nucleon pole term, denoted by $U^{\mu\nu}(p',s';p,s;q)$, which
is obtained by the insertion of two vertices ${\cal L}_1$. The latter can be expressed
through the nucleon vertex (an explicit expression is given below). 
The matching condition at $O(A^2)$ takes the form:
\begin{widetext}
\eq\nonumber \label{mA2}
&&\sum_{l,m,n=0}\,(-ip')_{i_1}\dots (-ip')_{i_n}(ip)_{j_1}\dots (ip)_{j_m}(iq)_{\mu_1}\dots (iq)_{\mu_l}\Pi_{s's}^{i_1\dots i_n,\,j_1\dots j_m,\,{\mu_1}\dots{\mu_l},\,\mu\nu}
\nonumber\\[2mm]
&=&T^{\mu\nu}(p',s';p,s;q)-U^{\mu\nu}(p',s';p,s;q)\, .
\en
\end{widetext}

As can be seen, the low-energy constants of the effective field theory are uniquely fixed 
by the Taylor  expansion of the nucleon vertex and the Compton scattering amplitude
in the external three-momenta.

\section{Energy shift}

Up to now, we have considered the problem in a generic external field. We next limit 
ourselves to a  static periodic magnetic field of the form
\eq
	{\bf B}=(0,0,B_3),\quad B_3=-eB\cos(\omega x_2), 
	\en	
where $B$ denotes the strength  of the field and the real parameter $\omega$ takes  nonzero values. 
The components of the gauge field $A^\mu(x)$ read
\eq\label{A-potential}
	A_1=\frac{eB}{\omega}\sin(\omega x_2),\qquad A_0=A_2=A_3=0.
	\en		
The parameter $\omega$ allows one to scan the virtuality of the photon. 

Since lattice simulations are performed in a finite spatial volume, the magnetic 
flux is quantized \cite{'tHooft:1979uj}. Consequently,  the parameter $\omega$ can 
take only  particular values (see, e.g., Refs. \cite{Davoudi:2015cba,Bali:2015msa}),
\eq\label{omega_n}
\omega=\frac{2\pi N}{L},\qquad N\in\mathbb{Z} \backslash \{0\},
\en
whereas the field strength $B$ is not quantized. 
Here, $L$ denotes the spatial size of the lattice. The quantization condition 
Eq.~(\ref{omega_n}) also guaranties the proper implementation of the magnetic 
field on a torus \cite{Davoudi:2015cba}. We note that, for $\omega\neq 0$, there
exists an alternative procedure that implies the quantization of the field strength $B$ 
instead of $\omega$~\cite{Davoudi:2015cba}. In the present paper, however, we do not 
consider this option.

The quantum-mechanical Hamiltonian, which acts on the 
single-nucleon wave function as a differential operator, can be straightforwardly derived from the 
effective Lagrangian, Eq.~(\ref{L_eff_A}).
It is convenient to first  rescale the nucleon field
\eq
\psi(\m x,t)\rightarrow\frac{1}{\sqrt{2W(\m\nabla)}}\,\psi(\m x,t).
\en
The terms in the Hamiltonian can be again ordered, according to the powers of $A_\mu$:
\eq
H=H_0+H_1+H_2+O(A^3).
\en
For instance, the free Hamiltonian reads
$H_0=W(\bs \nabla)\delta_{s's}\,$, and so on.
To the best of our knowledge, an analytic solution of the Schr\"odinger equation in the
periodic magnetic field is not available in the literature. For this reason, we resort to 
perturbation theory. The solutions must obey  periodic boundary conditions.
Denoting the Hilbert-state vector  in the nonrelativistic theory, 
corresponding to the unperturbed solution, by
$|{\bf k}_n, s\rrangle$, one has
\eq
\llangle{\bf x}|{\bf k}_n, s\rrangle=\frac{1}{L^{3/2}}e^{i\m k_n\m x}\chi_s\, .
\en
where $\chi_s$ denotes a Pauli spinor and
\eq
w(\m k_n)=\sqrt{m^2+\m k_n^2}, \qquad \m k_n=\frac{2\pi \bf n}{L}, \qquad \m n\in \mathbb{Z}^3.
\en
These vectors are normalized, according to
$\llangle {\bf k}_{m},s'|{\bf k}_n,s\rrangle 
=\delta_{{\bf m}{\bf n}}\delta_{s's}\, .$

Next, we  evaluate the energy shift of the nucleon
  ground state up to order $A^2$. Since the unperturbed solution
is twofold degenerate due to the nucleon spin, 
one should use perturbation theory for the degenerate 
states. First, one may check that the first-order  energy shift vanishes identically 
for $\omega\neq 0$, since, in this case,
$\llangle {\bf 0},s|H_1|{\bf 0},s'\rrangle=0$ due to three-momentum conservation. 
Further, in order
to determine the spin-averaged shift $\delta E$ at  second order, the calculation
of the diagonal matrix elements suffices:
\eq
\delta E=\frac{1}{2}\,\sum_s(\delta E_s'+\delta E_s{''})\, ,
\en
where the first contribution (the nucleon pole term)
emerges from the second iteration of $H_1$ and the second term is the matrix 
element of $H_2$. The explicit expressions are 
\eq
\delta E_s'&=&\sum_{\m k_n\neq\,\m 0}\sum_\sigma\,\frac{\llangle {\m 0,s}|H_1|{\m k_n, \sigma}\rrangle \llangle {\m k_n,\sigma}|H_1|{\m 0, s}\rrangle }{w(\m 0)-w(\m k_n)},
\nonumber\\[2mm]
\delta E_s{''}&=&\llangle {\bf 0},s|H_2|{\bf 0},s\rrangle\, .
\en
The calculation of these corrections is straightforward and we obtain
\eq
\delta E_s'&=&\frac{(eB)^2}{8m\omega^2}
[F(\bs \omega)+F(-\bs\omega)]\, ,
\nonumber\\[2mm]
\delta E_s{''}&=&-\frac{(eB)^2}{4m\omega^2}\left[  T^{11}(0,s;0,s;\hat q)-U^{11}(0,s;0,s;\hat q)\right]\, ,
\nonumber\\
\en
where $\bs\omega=(0,\omega,0)$, $\hat q=(0,\bs\omega)$ and
\eq
F(\bs \omega)=\sum_{\sigma}\frac{\langle \m 0,s |j^1(0)|\bs \omega, \sigma\rangle\langle \bs \omega, \sigma|j^1(0)|\m 0, s\rangle}{2w(\bs \omega)(w(\m 0)-w(\bs \omega))}.
\en
The crucial step in getting these formulas has been to apply the matching conditions 
Eqs.~(\ref{mA}) and (\ref{mA2}). Note also that here, having used the matching condition,
we switched back to the relativistic normalization of the state vectors.

It remains  to calculate the quantity $U^{11}(0,s;0,s;\hat q)$. 
Inserting a complete set of the one-nucleon states into the pertinent matrix element 
and using the matching condition Eq.~(\ref{mA}) once more, we obtain
\eq
U^{11}(0,s;0,s;\hat q)=-\frac{1}{2}\,[F(\bs \omega)+F(-\bs \omega)]\, .
\en
Finally, adding everything together, we get a remarkably simple formula
for the spin-averaged energy shift
\begin{eqnarray}\label{finalresult}
\delta E&=&\frac{(eB)^2}{4m}\,T_1(0,-\omega^2)+O(B^3)
\nonumber\\[2mm]
&=&\frac{(eB)^2}{4m}\,S_1(-\omega^2)+O(B^3)\, ,
\end{eqnarray}
where the expression for the spin-averaged tensor Eq.~(\ref{Taveraged}) 
has been explicitly used. This is the main result of the present work. Note that
the quantity $S_1$ here is the full one and not the inelastic part only.

\section{Discussion and conclusions}
The applicability of the formula Eq.~(\ref{finalresult}) is limited for several reasons. 
First, the strength of the magnetic field has to be chosen sufficiently small, 
so that  perturbation theory provides a meaningful result. A necessary condition for 
this is that the structure of the perturbed and the unperturbed spectrum remains the
same. 

An estimate for the upper bound on the magnitude of $B$ can be obtained
as follows. It can be straightforwardly checked that using the periodic external field,
 Eq.~(\ref{A-potential}), in the Schr\"odinger equation leads to a periodic potential
with  magnitude $V_0=e^2B^2/(2m\omega^2)$ and  period $d=\omega^{-1}$.
Considering a  single period as a potential well,  perturbation theory is 
applicable, if the well is shallow enough, so that no bound states are formed
(in the periodic potential, the band structure arises instead of the isolated 
energy levels).
Using, for simplicity, the known formula for the square well gives the condition
$eB<2\omega^2$. Of course, this should only be considered as a crude order-of-magnitude
estimate of the critical value of $B$. Note also that, from the point of view of 
phenomenological applications, the region of the small 
$\omega^2$ (smaller than a few $\mbox{GeV}^2$) is the most relevant one. 

As seen from Eq.~(\ref{finalresult}), the perturbative result is valid up to terms of 
order $B^3$. Albeit, in principle, it is possible to give some crude estimate of the
neglected terms by using ChPT, it is important to note that the validity of the formula
can be checked {\it a posteriori} on the lattice---by ensuring that the energy shift
grows quadratically with $B$.

On the other hand, the value of $B$ should be large enough so that the energy shift
$\delta E$ is measurable. Moreover, the accuracy of the extraction should be 
sufficient to allow for a disentanglement of the inelastic and elastic contributions. 
The latter is given by~\cite{Gasser:2015dwa}
\eq\label{eq:elastic}
S_1^{el}(q^2)=-\frac{4m^2}{q^2(4m^2-q^2)}\,\biggl\{G_E^2(q^2)-G_M^2(q^2)\biggr\}\, ,
\en
where $G_E,G_M$ denote the electric and magnetic form factors, respectively. 
Further, the inelastic contribution at $q^2=0$ is related to the magnetic 
polarizability by~\cite{Gasser:2015dwa}
\eq
S_1^{inel}(0)=-\frac{\kappa^2}{4m^2}-\frac{m}{\alpha}\,\beta_M\, ,
\en
where $\kappa$ denotes the anomalous magnetic moment of a nucleon and $\alpha$ 
is the fine-structure constant.
Experimental values for the proton and neutron polarizabilities are 
$\beta_M^p=(3.15\pm 0.50)\cdot 10^{-4}~\mbox{fm}^3$~\cite{McGovern:2012ew}
and $\beta_M^n=(3.65\pm 1.50)\cdot 10^{-4}~\mbox{fm}^3$~\cite{Myers:2014ace},
respectively. As already mentioned, little is known about the $q^2$ dependence
of $S_1(q^2)$. For a crude estimate, however, we assume that~\cite{Gasser:2015dwa}
\eq
S_1^{inel}(q^2)=\frac{S_1^{inel}(0)}{(1-q^2/0.71\,\mbox{GeV}^2)^2}\, .
\en
Using now Eq.~(\ref{finalresult}), it is immediately seen that the  inelastic
shift $\delta E^{inel}$ obeys the following relation:
\eq\label{eq:bound}
eB=\biggl(\frac{4m\,\delta E^{inel}}{S_1^{inel}(q^2)}\biggr)^{1/2}
=c\times |\delta E^{inel}|^{1/2}\, .
\en
Taking now $-q^2=\omega_{max}^2=2~\mbox{GeV}^2$ as the upper bound of the interval,
we get a crude estimate of the coefficient $c=2.97~\mbox{GeV}^{3/2}$ for the proton and
$c=2.77~\mbox{GeV}^{3/2}$ for the neutron. Note that, using a rather 
generous estimate $|\delta E^{inel}|=0.05\, m$,
the bound Eq.~(\ref{eq:bound})
at $-q^2=\omega^2$ is comfortably consistent
with the upper bound $eB<2\omega^2$, except very small values of $\omega^2$.

The total energy shift $\delta E=\delta E^{el}+\delta E^{inel}$ can be much 
larger than the inelastic shift alone, especially as $\omega^2$ is small. 
As seen from Eq.~(\ref{eq:elastic}), the elastic contribution is singular as $\omega^2\to 0$, whereas the inelastic contribution is regular. As an order-of-magnitude estimate,
one may ask, at what value of $\omega^2$ the magnitude of the inelastic
contribution amounts to a $10~\%$ of the singular piece of the elastic contribution, 
which behaves like $1/\omega^2$. The estimate gives 
$\omega_{min}^2=0.11~\mbox{GeV}^2$ for the proton and 
$\omega_{min}^2=0.05~\mbox{GeV}^2$ for the neutron.
Of course, setting a lower bound on the $\omega^2$-interval critically depends on our
 ability to extract the proton and neutron form factors on the lattice with high accuracy.
Note also that, even for the lowest value of $\omega_{min}^2$, the quantity
$M_\pi L$ is of order of 4---in other words, the calculations can be performed at the volumes which are feasible at present (here, $M_\pi$ is the pion mass).

Another interesting issue is the study of the limit $\omega\to 0$. In the
present framework, this limit is singular and is intertwined with the limit $L\to\infty$.
Indeed, recall that the values of $\omega$ are quantized: $\omega=2\pi N/L$. 
For nonzero values of $N$, the limit $\omega\to 0$ thus implies $L\to\infty$. In addition,
the Landau levels are bound even if $L\to\infty$, which violates the condition of the weak
$B$ field. We expect that the alternative setting for the periodic magnetic field on the
lattice (quantized $B$, $\omega$ not quantized), see Ref.~\cite{Davoudi:2015cba}, 
which has not been considered in the 
present paper, will be more advantageous for studying the limit $\omega\to0$.
Also, this different approach will allow one to continuously scan the interval 
of interest in the 
variable $q^2$. We plan to address these and other issues in a forthcoming
work~\cite{future}.

To summarize,  our final expression, Eq.~(\ref{finalresult}), enables one to directly
extract the subtraction function $S_1(q^2)$ from the lattice measurement of the nucleon
energy levels in a periodic external magnetic field.  

\begin{acknowledgments}
 \section{Acknowledgments} 
We thank Z.~Davoudi, J.~Gasser, H.~Leutwyler, J.~A.~Ol\-ler, M.~Savage, G. Schierholz
 and N.~Tantalo
 for useful discussions. 
We acknowledge the support from the DFG (CRC 110 ``Symmetries and the Emergence of Structure in QCD'' 
and  Bonn-Cologne Graduate School of Physics and
Astronomy). This research is supported in part by Volkswagenstiftung
under Contract No. 86260 and by the Chinese Academy of Sciences (CAS) 
President's
International Fellowship Initiative (PIFI) (Grant No. 2017VMA0025).
\end{acknowledgments}


\end{document}